\documentclass[%
    11pt,
    pdftex,
    normalheadings,
    DIV11,%
    parindent,
]{scrartcl}

\usepackage{colortbl}
\definecolor{darkred}{rgb}{0.6,0.0,0.0}
\definecolor{darkgreen}{rgb}{0.0,0.4,0.0}
\definecolor{darkblue}{rgb}{0.0,0.0,0.6}
\definecolor{boxgray}{rgb}{0.7,0.7,0.7}
\definecolor{identifierclr}{rgb}{0.286,0.454,0.752}
\definecolor{keywordclr}{rgb}{0,0,0}
\definecolor{stringclr}{rgb}{0.5,0.5,0.5}
\definecolor{backgroundclr}{rgb}{1,1,1}

\usepackage{pifont}

\usepackage{eulervm}
\usepackage{concrete}

\usepackage[automark]{scrpage2}

\usepackage[savemem]{listings}
\lstdefinelanguage{CSS}{
    sensitive=false,
    alsolanguage=HTML
}

\usepackage{graphicx}
\graphicspath{{images/}}

\usepackage{pspicture}
\usepackage{pstricks}
\usepackage{pst-all}
\usepackage{pst-blur} 
\usepackage{multido}

\usepackage{tabularx}
\newcolumntype{F}{>{\bfseries\sffamily}p{2.0cm}} 
\newcolumntype{S}{>{\footnotesize\sffamily}l} 

\usepackage[german]{varioref}

\usepackage{verbatim}

\usepackage{footnpag}

\usepackage{fancybox}

\usepackage[right]{eurosym}

\usepackage{bibgerm}

\usepackage{makeidx}

\usepackage[german, ngerman, english]{babel}
\usepackage{ae}

\usepackage[latin1]{inputenc}
\usepackage[T1]{fontenc}

\usepackage{array}

\usepackage{afterpage}

\usepackage{graphicx}
\usepackage{picins}

\addtokomafont{caption}{\small}
\setkomafont{captionlabel}{\sffamily\bfseries}
\setkomafont{footnote}{\sffamily}
\setkomafont{pagehead}{\normalfont\sffamily} 
\setkomafont{pagenumber}{\normalfont\sffamily}

\setcapindent{1em}

\usepackage{amsmath}
\usepackage{amssymb}

\usepackage{rotating}

\usepackage{color}

\definecolor{LinkColor}{rgb}{0,0,0.5}
\usepackage[%
    pdftitle={Submission of Content to a Digital Object Repository Using a Configurable Workflow System},%
    pdfauthor={Andreas Hense, Johannes M\"{u}ller},%
    pdfcreator={MiKTeX, LaTeX with hyperref and KOMA-Script},
    pdfsubject={},
    pdfkeywords={}]{hyperref}
\hypersetup{colorlinks=true,%
    linkcolor=LinkColor,%
    citecolor=LinkColor,%
    filecolor=LinkColor,%
    menucolor=LinkColor,%
    pagecolor=LinkColor,%
    urlcolor=LinkColor}

    {\begin{list}{$\diamondsuit$}{}}%
    {\end{list}}

\title{Submission of Content to a Digital Object
Repository Using a Configurable Workflow System}
\author{Andreas Hense, Johannes M\"{u}ller\\
University of Applied Sciences Bonn-Rhein-Sieg\\
Grantham-Allee 20\\
53757 Sankt-Augustin\\
Germany\\
\texttt{andreas.hense@fh-brs.de}\\
\texttt{johannes.mueller@gmx.com}}

\begin{document}

\maketitle

\begin{abstract}
The prototype of a workflow system for the submission
of content to a digital object repository is here presented.
It is based entirely on open-source standard components and features a service-oriented architecture.
The front-end consists of Java Business Process Management (jBPM), Java Server Faces (JSF), and Java Server Pages (JSP).
A Fedora Repository and a mySQL data base management system serve as a back-end.
The communication between front-end and back-end uses a SOAP minimal binding stub.

We describe the design principles and the construction of the prototype and discuss the possibilities and limitations of workflow creation by administrators.
The code of the prototype is open-source and can be retrieved in the project \emph{escipub} at \url{http://sourceforge.net}.
\end{abstract}

\section{Motivation}

This work has been inspired by the eSciDoc project of the Max-Planck-Society \cite{Q01_v1}. One of the goals of the eSciDoc project is the creation of a publication management service that allows scientific organizations to establish an institutional repository.

Generally speaking, the publication process goes like this.
Publications, consisting of a set of metadata and a number of content files, are submitted to a digital repository and are made publicly available following the philosophy of open access. Once publications are available they can be retrieved by a so-called persistent identifier. The organization that uses persistent identifiers guarantees that publications can always and forever be retrieved by that identifier.
The responsibility for the quality of academic output, the cost of long term archival, and the fact that publications cannot be withdrawn make a well-structured quality assurance process almost mandatory.

The quality assurance process is a business process that must be adaptable to the specific needs of the organization using it.
This goes beyond the mere assignment of roles to actual persons.
The process must be configurable with respect to the number of steps (e.\,g.\ technical, formal, and scientific quality assurance),
the degree of parallelization of different tasks, and even the quality of the tasks themselves.

The business processes of submission and quality assurance are highly structured and can be automated \cite{Q71}. The part of a business process that is automated is called a workflow.

As a consequence, the software-architecture of choice for the submission of content to an institutional repository consists essentially of two parts: a workflow system and a digital object repository.

\section{Features of the Prototype}
The prototype is based on the \emph{jBPM Starters Kit 3.1.1} \cite{Q152} by JBoss, which has been extended by an interface to Fedora. 
The prototype is called "eSciPub" and can be tested under the following URL:\\ \url{http://www.bis.inf.fh-brs.de}.

The prototype features a simple role model. Users log into the system with their account and can perform tasks according to their role (author, quality assurer, 
process administrator,...). Every role has its proper workspace.
The workspace of an author shows his pending submissions, his submissions that came back from quality assurance for rework, and a list of publication processes.
The author can submit a new article by choosing one of theses processes.

Every time a user is working on a task of a workflow instance, its state instance is shown in a diagram. The diagram lists the whole workflow with the current task being highlighted
(cf. figure \ref{fig:screenshot}).

The administration of workflow instances like advancing and stopping can be performed by the process administrator using the same software as the other users.
New workflows or new versions of existing ones are created graphically in the Eclipse development environment. They can also be deployed from there. Note that running workflow instances can continue with their original definition.

\section{The Software Architecture}
The prototype presented in this paper is built on open-source components and is itself open-source.
\subsection{JBoss jBPM}
\label{sec:JBossjBPM}

jBPM \cite{jbpm} is a workflow management system by JBoss that is implemented in Java.
jBPM has its roots in an open-source project initiated by Tom Baeyens in 2003 and which has been managed by JBoss since 2004.

The central component of jBPM is a workflow engine running processes described in XML.
jBPM supports jPDL (\emph{jBPM Process Definition Language}) and BPEL (\emph{Business Process Execution Language}).
We have chosen jPDL because BPEL was added recently as the so-called ``JBoss jBPM BPEL Extension'' and is still in the beta testing phase.

All relevant data for a process are stored in a single compressed file -- a process archive.
The process definition itself is a file in the process archive called \texttt{processdefinition.xml}.
If a process definition is created by the \emph{Graphical Process Designer} (GPD), which we will describe below, a picture of the process called
\texttt{processimage.jpg} and a file containing metadata of the picture called  \texttt{gpd.xml} are added to the process archive.
All Java classes and libraries used by the process and all documents used as information by process owners can be added to the process archive.

Process definitions are directed graphs consisting of nodes an edges. The edges are called transitions.
Nodes have a type determining their properties and behavior \cite{Q21_v1}. 
Every process definition has a name. XML-editors like GPD can check the validity of the jPDL-XML-structure using the 
XML-Namespace  \texttt{urn:jbpm.org:jpdl-3.1}.

Tasks are mapped to users or roles by the jPDL-construct \emph{swimlane}.
When the first task in a swimlane is created, the \emph{assignment handler} of this swimlane is called;
the assignment handler is defined by the attribute \texttt{assignment} of the process definition.
The class referenced by the assignment handler could serve for user authentication purposes.

Process variables can be simple data types or complex Java-objects -- if they are serializable.
Process variables can be contained in the process definition or can be created during process execution.

\emph{Actions} contain a programming logic that is executed when a certain event happens during process execution.
Typical events triggering actions are arriving at a node, leaving a node, or using a transition.
Actions assigned to general events and not to nodes cannot influence the control flow of the process instance.
In contrast, actions assigned to nodes are responsible for the control flow of the process instance \cite{Q21_v1}.

\begin{figure}[!htbp]
\centering \includegraphics[width=\textwidth]{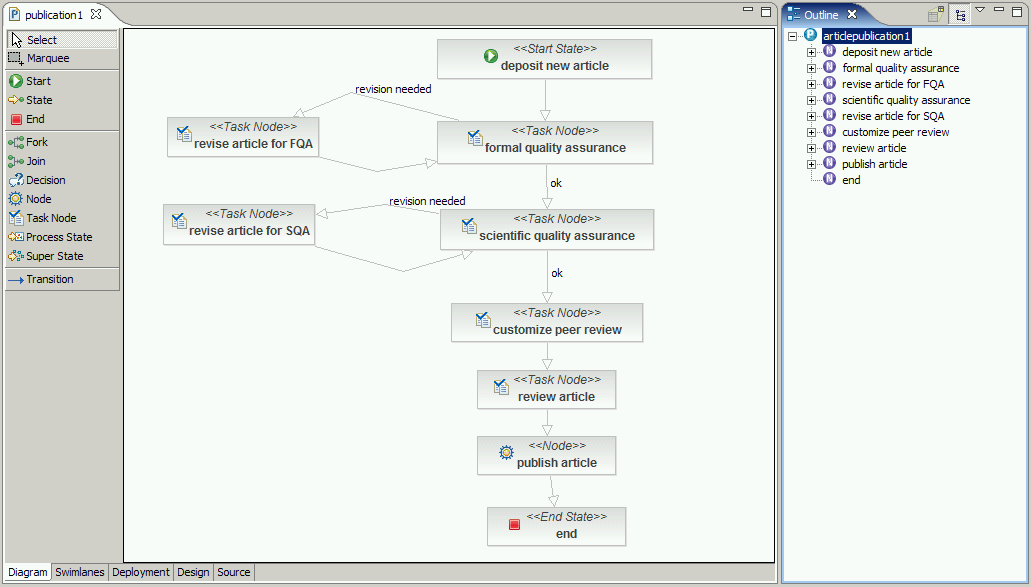}
\caption{jBPM Graphical Process Designer (GPD).}
\label{fig:gpd}
\end{figure}

Process definitions can be created by the \emph{Graphical Process Designer} \cite{Q152}, an Eclipse-plugin (cf. figure \ref{fig:gpd}).
Using GPD, processes can be designed graphically and be deployed on a running jBPM-server.
Older versions of process definitions are not overwritten and running process instances of the older versions can be terminated using their original definition.

\subsection{Fedora}
\label{sec:FedoraRepository}

The \emph{Fedora Repository}\footnote{Fedora stands for ``Flexible Extensible Digital Object Repository'', cf.~{\url{http://www.fedora.info}}}
stores and manages digital objects. Fedora was selected by the eSciDoc-project as the basic technology for the institutional repository of the Max-Planck-Society \cite{Q91}.

A \emph{Fedora-object} consists of a collection of so-called ``\emph{Content Items}'' and associated metadata.
Content items contain documents, images, or any other type of file.
An important feature of Fedora is that so-called ``\emph{disseminators}'' can be associated with content items.
This feature is essential for any scientific document server:
the original \LaTeX\footnote{Type setting system \emph{\LaTeX}, s.~{\url{http://www.latex-project.org/}}}-file
of this article could be contained in a content item of a Fedora-object. While the original file is only visible to the author, disseminations 
in HTML- or pdf-format could be presented to the public.
In a slightly different context, original high definition images could be contained in content items and low resolution copies could be made publicly available by a disseminator.

Fedora is prepared for operation in a service-oriented architecture. Via the SOAP-interfaces \texttt{API-A} and \texttt{API-M},
content can be stored and retrieved in various ways. Fedora comes with an integrated Tomcat-webserver making its content accessible via HTTP.
Content items can be stored locally or as references to other systems. Versioning of Fedora-objects is provided.

\subsection{Further Technologies}
The user interface is implemented using \emph{Java Server Faces} (JSF) (\emph{MyFaces} cf.~{\url{http://myfaces.apache.org}}).
JSF is a framework by Sun for the implementation of web applications. 
MyFaces is the first open-source implementation of JSF.
JSF is made for processing user interactions. Its interfaces are made of elements having a state. The states of elements and events can be supervised by the JSF-instance. The tag libraries of JSF can be used in Java Server Pages (JSP).
JSF runs as a servlet on the Tomcat servlet container.

The open-source data base management system \emph{MySQL}\footnote{cf.~\url{http://www.mysql.com}} is used for \emph{JBoss jBPM} and the \emph{Fedora Repository}.

For accessing the SOAP-interface, the Apache \emph{Axis}-library (Apache eXtensible Interaction System, cf.~\url{http://ws.apache.org/axis/}) is used.
Axis is a SOAP-engine for the construction of web services and clients.

XML-documents are constructed and accessed with the \emph{Document Object Model} (DOM)-library of the \emph{World Wide Web Consortium} (W3C)
(cf.~\url{http://www.w3.org/DOM/}).

The component library Apache \emph{Tomahawk} is an extension of the MyFaces-implementation and is used for making
Java Bean attributes persistent (cf.~\url{http://myfaces.apache.org/tomahawk/index.html}).

\subsection{Layers}

\begin{figure}[!htbp]
\centering \fbox{\includegraphics[width=0.9\textwidth]{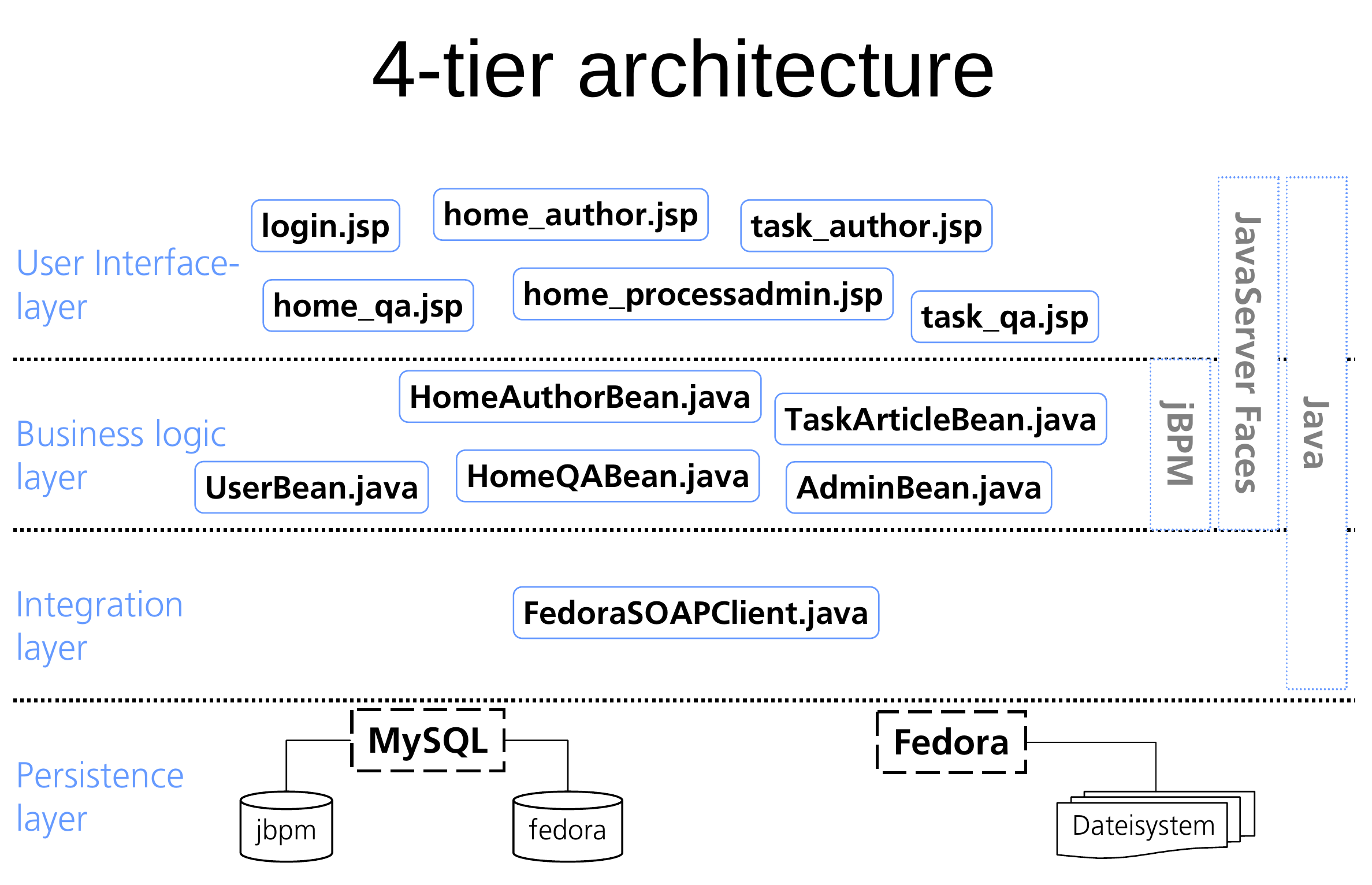}}
\caption{4-tier architecture of the prototype.}
\label{fig:4tier}
\end{figure}

The prototype is divided into four layers. Program modules of one layer only communi\-cate with program modules of adjacent layers
(cf.~figure~\ref{fig:4tier}). The layers are as follows:
\begin{enumerate}
  \item \emph{User-Interface layer:} contains the JSP/JSF-pages for the interaction with the user.
  \item \emph{Business-logic layer:} contains the \emph{backing beans} and Java objects 
implementing the ``business logic''.\footnote{This layer could be further divided into a layer of backing beans for the JSP and a business logic layer.}
  \item \emph{Integration layer:} Objects in this layer communicate with the data stores. Their only task is taking requests from the business-logic layer
and passing them to the data store -- or vice versa.
  \item \emph{Persistence layer:} This layer contains no Java objects, but the MySql DBMS and the Fedora Repository.
\end{enumerate}

\section{Accessing Fedora's Web Interface}

\subsection{Choosing the Right Way}

There are several ways of accessing Fedora via its interfaces.
Fedora offers a REST and a SOAP interface for calling the functions of \texttt{API-A} and \texttt{API-M}.
Being the more popular protocol, we have chosen SOAP for our prototype \cite{Q80}.
Once the decision for SOAP is made, there are the following possibilities:

\begin{enumerate}
  \item Using the \emph{SOAP-Client} of the Fedora distribution \cite{Q143}.
This HTTP-Client is restricted to \texttt{API-A}-calls.
For the functionality of our prototype, \texttt{API-M} is the more important interface.
A first approach to using the program logic of the official Fedora client and extending it to \texttt{API-M}-calls
was abandoned, because the Fedora client is too tightly interwoven with a number of ``heavy'' classes of its distribution.
  \item Another approach is the direct access to Fedora using SOAP. Here, there are several sub-variants.

The variant implicitly suggested by the official Fedora documentation is the use of the Fedora client library ``\texttt{client.jar}''.
This library has a size of nearly one MB in version 2.1.1.
It provides the data types necessary for SOAP requests to Fedora -- e.\,g.\ \\
\texttt{fedora.server.types.gen.Datastream}\\
or \texttt{fedora.server.types.gen.RepositoryInfo}

We have chosen a more elegant - although not officially documented method: we have used a so-called \emph{minimal binding stub} that can 
be generated automatically (cf. next section). Its size is 89 KB when packed as a jar-file.
\end{enumerate}

\noindent Our approach follows the principle of \emph{loose coupling} in a service-oriented architecture, because the client is independent of the ``heavy'' server-specific class libraries. It creates the necessary classes itself. This approach is less resource-hungry and results in a higher performance.

\subsection{Generating the SOAP-Binding Stub}

If programs want to use the SOAP-interface of the Fedora server, the generic data types of Fedora must be known in the runtime environment of the client program.
To achieve this, there are two possibilities: include all Java classes of the Fedora implementation as source files or a jar-file, or include a \emph{minimal binding stub}. Such a binding stub contains only those data types that the client needs to send to or receive from the web-service-provider. As already pointed out, the last possibility is the one we prefer.

A minimal binding stub can be generated by analyzing the description of the web service.
This is done by the tool ``\texttt{WSDL2Java}'' from the \emph{Apache Axis} framework.
``\texttt{WSDL2Java}'' analyses WSDL-files (\emph{Web Services Description Language}).
WSDL-files define the signatures, i.\,e.\ the names and parameters of web services.
In general, the non-primitive data types are defined in XSD-files (\emph{XML Schema Definition}) referenced by the WSDL-file.

On the basis of WSDL- and corresponding XSD-files, \texttt{WSDL2JAVA} builds skeleton Java classes.
The structure of the skeleton Java classes corresponds to the WSDL- and XSD-files.
The skeleton Java classes contain no program logic -- apart from simple getter and setter methods.
Thus the client is able to call the web services and process the answers correctly.

The WSDL-files for the \texttt{API-A}- and \texttt{API-M}-SOAP-interfaces of Fedora (cf.~section~\ref{sec:FedoraRepository}) can be found here:
\begin{itemize}
\item \url{http://www.fedora.info/definitions/1/0/api/Fedora-API-A.wsdl},
\item \url{http://www.fedora.info/definitions/1/0/api/Fedora-API-M.wsdl}.
\end{itemize}
If Fedora is installed on a standard port of the local host, the files can be found here:
\begin{itemize}
\item \url{http://localhost:8180/fedora/services/access?wsdl},
\item \url{http://localhost:8180/fedora/services/management?wsdl}.
\end{itemize}
Fedora-WSDL-files reference the XSD-file ( \emph{XML Schema Definition})\\
\url{http://www.fedora.info/definitions/1/0/types/fedora-types.xsd}.

 \texttt{WSDL2JAVA} needs the following class libraries for the creation of Java classes from Fedora-WSDL-files:
\begin{enumerate}
  \item All program libraries of the Axis distribution, because \texttt{WSDL2Java} is a part of Axis.
  \item With the \emph{JavaBeans Activation Framework}\footnote{cf.~\url{http://java.sun.com/products/javabeans/jaf/}}, unknown data types can be determined,
encapsulated, and their methods can be detected.
The only jar-file of this distribution called \texttt{activation.jar} is needed.
  \item The \emph{JavaMail API}\footnote{cf.~\url{http://java.sun.com/products/javamail/}} is a programming interface containing a protocol independent
framework for eMails and messages. The distribution contains several jar-files, of which only \texttt{mail.jar} is needed.
\end{enumerate}
The XML schema definition files \texttt{fedora-types.xsd} and\\
 \texttt{fedora-auditing.xsd} 
must be in the path of the operating system shell in use.
These files can be retrieved in the folder \texttt{xsd} or via the addresses 
\begin{itemize}
\item \url{http://localhost:8180/fedora-types.xsd} or 
\item \url{http://localhost:8180/fedora-auditing.xsd}
\end{itemize}
of a local Fedora-server installation\footnote{The URL referenced in the WSDL of the 
 \texttt{API-M}-interface  (\url{http://www.fedora.info/definitions/1/0/auditing/fedora-auditing.xsd})
was not available at the time of writing this document.}.
If all the WSDL-files, XML-schema-files, and class libraries mentioned above are available, the binding stub can be generated by using the following commands:
\begin{verbatim}
java -cp C:\java\axis-1_4\lib\axis.jar;
         C:\java\axis-1_4\lib\commons-logging-1.0.4.jar;
         C:\java\axis-1_4\lib\commons-discovery-0.2.jar;
         C:\java\axis-1_4\lib\jaxrpc.jar;
         C:\java\axis-1_4\lib\saaj.jar;
         C:\java\axis-1_4\lib\wsdl4j-1.5.1.jar
         C:\jaf-1.1\activation.jar;
         C:\javamail-1.4\mail.jar
         org.apache.axis.wsdl.WSDL2Java Fedora-API-A.wsdl

java -cp C:\java\axis-1_4\lib\axis.jar;
         C:\java\axis-1_4\lib\commons-logging-1.0.4.jar;
         C:\java\axis-1_4\lib\commons-discovery-0.2.jar;
         C:\java\axis-1_4\lib\jaxrpc.jar;
         C:\java\axis-1_4\lib\saaj.jar;
         C:\java\axis-1_4\lib\wsdl4j-1.5.1.jar
         C:\jaf-1.1\activation.jar;
         C:\javamail-1.4\mail.jar
         org.apache.axis.wsdl.WSDL2Java Fedora-API-M.wsdl
\end{verbatim}
At first, the generated structure of folders contains only Java source files.
These have to be compiled and be put into the class path of the project as a jar-archive.

\section{The Nuts and Bolts of the Web Application}
One of the roles in our submission process is that of the author. He submits new content to the digital object repository.
The workspace of the author (\texttt{home\_author.jsp}) contains three areas: ``Task-List'', ``Start New Publication Process'', and 
an overview of all articles of this author in the repository (cf. figure \ref{fig:screenshot}).
This section describes the mechanisms for addressing Fedora in the context of jBPM and JSF.

\subsection{Display the jBPM Task List and Execute a Task}
The JSF-page \texttt{home\_author.jsp} with its \texttt{dataTable}-tag represents a table that is filled with the data
from the \texttt{getTaskInstances}-method of the ``HomeAuthorBean''\footnote{\texttt{HomeAuthorBean.java}}:
\vspace{-1em}\begin{verbatim}
<h:dataTable value="#{homeAuthorBean.taskInstances}">
\end{verbatim}\vspace{-1em}This bean is administered by JSF and therefore need not be included in the JSP explicitly.
It has the scope ``\emph{Request}'' meaning that this bean is initialized for each request.
The \texttt{JbpmContextFilter} and the constructor of the HomeAuthorBean ensure that the correct user- and jBPM-context-information
is contained in the bean when the method is called by\\
 \texttt{home\_author.jsp}.

Using the class \texttt{org.jbpm.db.TaskMgmtSession}, the function\\
 \texttt{TaskAuthorBean.getTaskInstances} can access
the method \texttt{findTaskInstances}, which returns all open tasks of an actor, directly:
\vspace{-1em}\begin{verbatim}
taskMgmtSession.findTaskInstances(userBean.getUserName());
\end{verbatim}\vspace{-1em}
Using the column ``Name'' of the task list, which contains the names of the tasks as links, the function \texttt{selectTaskInstance} of the HomeAuthorBean can be called:\\
\verb|<h:commandLink action="#{homeAuthorBean.selectTaskInstance}">|

The author can submit a new article or work on an article already in the process of publication.
Both of these tasks are handled by the same backing bean ``TaskAuthorBean'' (\texttt{TaskAuthorBean.java}).
If the task comes from the task list, then it already exists in jBPM and does not have to be created.
In this case, the \texttt{selectTaskInstance}-method of the Home\-Author\-Bean hands over the unique identifier of the selected task - the  ``\emph{task\-InstanceId}'' -
to the ``TaskAuthorBean'' (\texttt{TaskAuthorBean.java}) and displays the next page \texttt{task\_author.jsp}:\\
\verb|return "task";|

This last JSP is backed by the TaskAuthorBean (cf. ``Submission of a new article'' and ``Work on an article'').

\subsection{Creation of a new Fedora object}
\label{sec:FedoraObjectCreation}

The table of process definitions on  \texttt{home\_author.jsp} is filled by the\\
 \texttt{getlatestProcessDefinitions}-method of the HomeAuthorBean:\\
\verb|<h:dataTable value="#{homeAuthorBean.latestProcessDefinitions}">|

This method calls the method \texttt{findLatestProcessDefinitions} of the jBPM-class\\
\texttt{org.jbpm.db.GraphSession} which returns a list of current versions of process definitions.
When a new publication process is started, the method \texttt{startProcessInstance} of the HomeAuthorBean is called.
This method creates a new Fedora object:\\
\verb|pid = fedoraSOAPClient.ingestNewFedoraObject();|

The object creation by the ``FedoraSOAPClient'' (class \texttt{FedoraSOAPClient}) is as follows:
\begin{enumerate}
  \item At first a new DOM-document is created as an instance of the object\\
  \texttt{org.w3c.dom.Document}:\\
\verb|Document foxmlDoc = docBuilder.newDocument();|
  \item The FOXML-root node, additional root node attributes, and the object properties \emph{Label} and \emph{Content Model} are added.
The FOXML-DOM-document now contains the structural information necessary for insertion into the repository.
  \item In order to hand over the DOM-document to Fedora by a SOAP-call, it must be serialized in a \emph{Byte Array} first.
Using the three statements
\begin{verbatim}
ByteArrayOutputStream xml = new ByteArrayOutputStream();
XMLSerializer ser = new XMLSerializer(xml, null);
ser.serialize(foxmlDoc);
\end{verbatim}
the DOM-document \texttt{foxmlDoc} is transformed to a\\
 \texttt{java.io.ByteArrayOutputStream}. 
  \item The SOAP-call is prepared by creating an instance of the \emph{Axis}-object\\
  \texttt{org.apache.axis.client.Service}:\\
  \verb|Service service = new Service();|
  \item Using the \texttt{service}-object, the SOAP-client can create an instance of\\
  \texttt{org.apache.axis.client.Call}. The latter calls a SOAP-remote procedure (RPC):\\
\verb|Call call = (Call) service.createCall();|
  \item Some parameters of the SOAP-client are set:
\begin{verbatim}
call.setOperationName(new QName(
  "http://www.fedora.info/definitions/1/0/api/",
  "ingest"));
call.setTargetEndpointAddress(new URL(
"http://localhost:8180/fedora/services/management"));
call.setUsername(FEDORA_SERVER_USERNAME);
call.setPassword(FEDORA_SERVER_PASSWORD);
\end{verbatim}
The newly created object of type \texttt{javax.xml.namespace.QName.QName} represents a \emph{Qualified Name},
which is connected to the namespace-URI of the Fedora-API. 
This qualified name contains the names of the SOAP-operation (``\texttt{ingest}'').
By using methods
\texttt{setTarget\-EndpointAddress} and \texttt{set\-Username}
the service-endpoint of the Fedora server and the credentials for authentification are set. The call is now finished.
  \item The \texttt{ingest}-method of the Fedora-API-M-SOAP-interface, besides the serialized FOXML-document, needs two more parameters~-- 
the XML-format of the object handed over and a protocol entry (\emph{Log Message}).
These are handed over during the execution of the \texttt{Call.invoke}-method:
\begin{verbatim}
String format = "foxml1.0";
String logmsg = "initial creation";
pid = (String) call.invoke(new Object[] {
  xml.toByteArray(),
  format,
  logmsg });
\end{verbatim}
  \item If the creation of the new object was successful, Fedora returns the automatically generated persistent identifier (PID) of the new object. By the entry\\
\verb|<param name="pidNamespace" value="escipub"/>|\\
in Fedora's configuration file\footnote{The Fedora configuration file can be found in server/config/fedora.fcfg} 
the PID-prefix has been set to ``escipub''.
This prefix is followed by a colon and a unique number. The answer to the FedoraSOAPClient will be e.\,g.\  ``escipub:477''.
\end{enumerate}

\noindent After the creation of a new Fedora-object, the control flow returns to the method\\
 \texttt{startProcessInstance} of the HomeAuthorBean.

If the creation of the Fedora-object was successful, a new process instance of the previously chosen process definition is created:
\begin{verbatim}
ProcessDefinition processDefinition =
  graphSession.loadProcessDefinition(processDefinitionId);
ProcessInstance processInstance =
  new ProcessInstance(processDefinition);
\end{verbatim}
Every process instance has a unique initial state.
By using the call
\begin{verbatim}
TaskInstance taskInstance =
  processInstance.getTaskMgmtInstance().createStartTaskInstance();
\end{verbatim}
the task corresponding to this initial state is created.
The \texttt{AuthenticationFilter}, the \texttt{JbpmContextFilter}, and the assignment of the \texttt{ActorId} in the jBPM-context make the new task to be assigned to the right actor and the
corresponding task list.
The PID is saved in the process context and is therefore available to all process participants as a process variable.
To make the process operations persistent, the jBPM-context is saved:
\begin{verbatim}
taskInstance.setVariable("pid", pid);
jbpmContext.save(processInstance);
\end{verbatim}

\subsection{Execution of a Fedora query}
We will sketch the steps that are necessary to get the list of the author's articles by a query to Fedora:

\begin{enumerate}

  \item The \verb|<h:dataTable>|-tag on \texttt{home\_author.jsp} expects from the HomeAuthorBean a list of articles of the current author.
Although this list is untyped at this level, it will be possible to access all attributes of the objects in the list, as long as
\emph{Getter}-methods are available in the Java context.

  \item The HomeAuthorBean formulates a query to the integration layer by specifying the maximum number of hits (100),
the comparison operator to use\\
\verb|info.fedora.www.definitions._1._0.types.ComparisonOperator|\footnote{This and all other Fedora types are described in \cite{Q151}.},\\
the field the query refers to (``creator''), and the value to check (the name of the current user). This query is handed over to the FedoraSOAPClient.

The result of the query to the integration layer is an object of type\\
\verb|info.fedora.www.definitions._1._0.types.FieldSearchResult|.\\
This type encapsulates the abstract type ``\texttt{resultList}'', which is of the (concrete) type \texttt{ArrayOfObjectFields}.
The attributes of an \texttt{ObjectFields}-object contain Dublin Core metadata like ``creator'', ``subject'', and ``description'', and
Fedora object properties like the PID or the creation date (``cDate'') \cite{Q151}.

  \item The method \texttt{doQuery} of the FedoraSOAPClient transforms the query coming from the HomeAuthorBean into an object of type\\
\verb|info.fedora.www.definitions._1._0.types.FieldSearchQuery|.\\
A \texttt{FieldSearchQuery} consists mainly of an array of conditions; thus queries with an arbitrary number of conditions can be handled.
In this case, we use only one condition.
The FieldSearchQuery is handed over to the method \texttt{findObjects}.

  \item In method \texttt{findObjects}, there is a SOAP call to the Fedora server as described above (section \ref{sec:FedoraObjectCreation}).
But this time, there are Fedora-specific data types that are unknown to the \emph{Axis}-library.
Thus, all Fedora data types of this SOAP-call are introduced to the Axis-client as qualified name objects before the \texttt{call.invoke}-statement
by the method 
\texttt{call.registerTypeMapping}, like for instance the data type \texttt{FieldSearchResult}\footnote{The classes \texttt{org.apache.axis.encoding.ser.BeanSerializerFactory} and\\ \texttt{org.apache.axis.encoding.ser.BeanDeserializerFactory} provide objects for the serialization and deserialization of new data types.}:
\begin{verbatim}
QName qn1 = new QName(
  "http://www.fedora.info/definitions/1/0/types/",
  "FieldSearchResult");
call.registerTypeMapping(
  FieldSearchResult.class,
  qn1,
  new BeanSerializerFactory(FieldSearchResult.class, qn1),
  new BeanDeserializerFactory(FieldSearchResult.class, qn1));
\end{verbatim}

  \item The answer of the server is of type FieldSearchresult. It is returned to the method \texttt{doQuery}.

  \item The \texttt{doQuery}-method hands on the answer to HomeAuthorBean.

  \item Before HomeAuthorBean passes on the information from the integration layer to the user-interface layer the monolithic
\texttt{FieldSearchResult}-object is transformed to a list of ObjectFields.
\texttt{home\_author.jsp} can access the entries of this list directly. 
The indexing shows that some of the Dublin Core attributes are arrays.
Indeed, the Dublin Core standard has repeatable attributes.
\end{enumerate}

\subsection{File Upload and Updating a Fedora Object}

\begin{figure}[p]
\centering \includegraphics[height=\textheight]{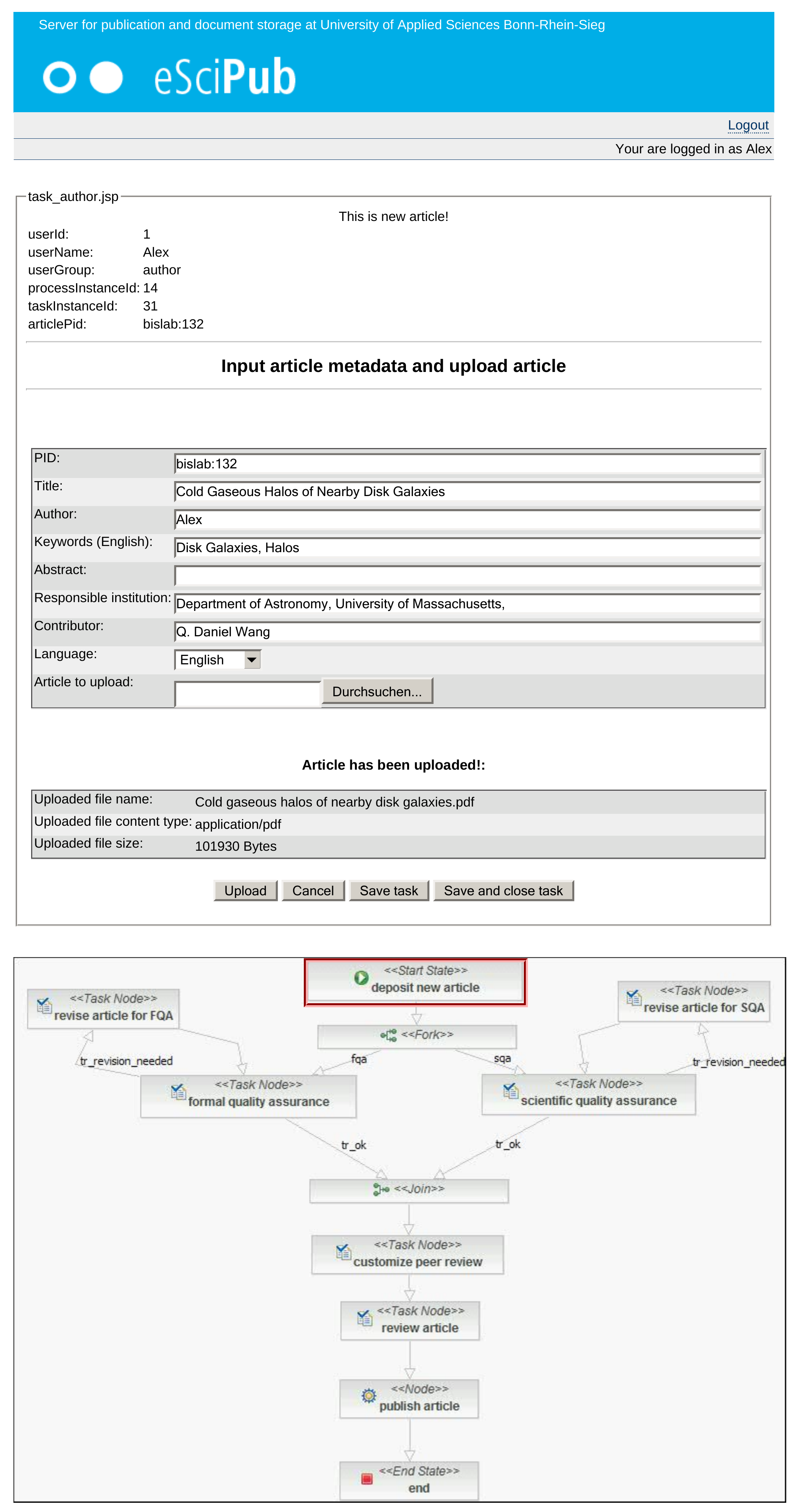}
\caption{Screenshot of the prototype - Input metadata and upload file}
\label{fig:screenshot}
\end{figure}

When the author sees the screen on page~\pageref{fig:screenshot}, a new empty Fedora object has already been created.
The PID field has been filled automatically by the system.
The author fills in the other metadata fields. Then he may want to upload a file.
We will now describe the file upload functionality, because this is a feature that is not contained in jBPM.

By the statement\\
\verb|<%@ taglib uri="http://myfaces.apache.org/tomahawk" prefix="t"%>|\\
in \texttt{task\_author.jsp} the \emph{Tomahawk}-tag-library is made available on this page.
By using the tag
\begin{verbatim}
<t:inputFileUpload id="fileupload"
                   accept="*/*"
                   value="#{taskArticleBean.upFile}"
                   storage="file" />
\end{verbatim}
a file selection dialogue is displayed.
The selected file of this dialogue is directly linked to the \texttt{upFile}-attribute of the TaskArticleBean.
This attribute is of type\\
\verb|org.apache.myfaces.custom.fileupload.UploadedFile| \\
from the \emph{Tomahawk}-library (\texttt{tomahawk.jar}).
Note that the file selection dialogue must be inside an HTML-form with attribute
 \verb|enctype="multipart/form-data"| 
because the \texttt{ExtensionsFilter} responsible for the upload in the JSF-context will not recognize it otherwise.
The \emph{upload}-button is linked to the \texttt{upload}-method of the TaskArticleBean:\\
\verb|<h:commandButton action="#{taskArticleBean.upload}" />|

\noindent By pressing the upload button on \texttt{task\_author.jsp}, the upload method of the TaskArticleBean is executed and a new HTTP-\emph{Request} is triggered.
Since the \emph{Scope} of the TaskAuthorBean is ``Request'',
this bean would be created and initialized as new by the JSF-servlet.
This means that the bean's attributes would be lost.
In order not to lose this information, all attributes of the TaskAuthorBean that must survive the end of a request are listed in
 \texttt{task\_author.jsp} via
\emph{Tomahawk}-tags \verb|<t:saveState...>|.

A remarkable feature of Fedora is the fact that files cannot be imported from the local file system but only via a web-URL.
That is why the file chosen for upload is first copied to the root directory of Fedora's \emph{Tomcat}-container.
From there it can be imported by the  \texttt{upload}-method of the TaskArticleBean using a local URL. 
The TaskArticleBean also provides important file information discovered while uploading, as for instance the \emph{content type} and the file size.
After uploading the article, the page \texttt{task\_author.jsp} is rebuilt, and the information provided by the TaskArticleBean about the uploaded file is displayed
(cf. figure \ref{fig:screenshot}).
A copy of the file now resides in the local \emph{Tomcat}-root-directory, e.\,g.\ in\\
\verb|U:/Master/fedora-2.1.1/server/jakarta-tomcat-5.0.28/webapps/ROOT/|

\noindent When the author closes the task by pressing the button ``Save task and quit'',
the method \texttt{saveAndCloseAuthor} of the TaskArticleBean is called.
This method saves the metadata of the form on \texttt{task\_author.jsp} in jBPM-process variables,
so that other roles involved in the same process, e.\,g.\ the quality assurance, need not get these metadata from Fedora, but
can access these process variables directly.

After that, the TaskArticleBean saves the metadata in the corresponding Fedora object.
The PID for accessing the correct Fedora object can be read from the process variable and be handed over to the FedoraSOAPClient:
\begin{verbatim}
boolean success = fedoraSOAPClient.changeDC(
  articlePid, userBean.getUserName(),
  articleTitle, articleCreator, articleSubject, 
  articleDescription, articlePublisher, 
  articleContributor, articleDate, articleType,
  articleLanguage, articleCoverage, articleRights);
\end{verbatim}

\noindent The method \texttt{changeDC} of the FedoraSOAPClient can change the metadata. 
Here, the new Dublin Core-data stream is built as a DOM-document: 
at first a new DOM-document is created with the necessary Dublin Core-namespace-attributes.
Then the DC-metadata are inserted as additional nodes according to the DC-namespace-specification\footnote{s.~\url{http://purl.org/dc/elements/1.1/}}.
Since Fedora creates a DC-data stream for each new object automatically, the FedoraSOAPClient uses the \texttt{API-M}-method \texttt{modifyDatastreamByValue} to save
the metadata in Fedora:
\begin{verbatim}
call.setOperationName(new QName(
  "http://www.fedora.info/definitions/1/0/api/",
  "modifyDatastreamByValue"));
\end{verbatim}

\noindent The new DOM-document containing the Dublin Core metadata is transformed to a \texttt{Byte Array} and handed over to Fedora:
\begin{verbatim}
call.invoke(new Object[] { pid, // Die PID
  "DC", // name of the data stream
  null, // altIds (no alternative identifier)
  null, // dsLabel (no change, still "ESCIPUB")
  new Boolean(true), // versioning on
  "text/xml", // MIME Type: Text
  null, // formatURI (none)
  xml.toByteArray(), // dsContent (the serialized Dublin Core)
  "A", // dsState (state - active)
  "update", // logMessage (log-entry)
  new Boolean(true) }); // force
\end{verbatim}

Using the method dsExists, the FedoraSOAPClient has the TaskArticleBean find out, if the data stream with the label ``ARTICLE'' exists. 
This check is necessary because the TaskArticleBean is also used for reworking an existing article.
Prior to saving the article in Fedora, the MIME type of the uploaded file in the local \emph{Tomcat}-root-directory is detected:
\begin{enumerate}
  \item By the statements
\begin{verbatim}
MultiThreadedHttpConnectionManager cManager =
  new MultiThreadedHttpConnectionManager();
HttpClient httpClient = new HttpClient(cManager);
\end{verbatim}
a new HTTP-client of type\\
 \texttt{org.apache.commons.httpclient.HttpClient} is created.
  \item The URL of the uploaded file is contained in the string variable \texttt{localURL}. The call
\begin{verbatim}
org.apache.commons.httpclient.methods.HeadMethod head =
  new HeadMethod(localURL);
\end{verbatim}
triggers the MIME-type detection  of this file.
\end{enumerate}

\noindent If the ARTICLE-Datastream exists already in the corresponding Fedora object, the FedoraSOAPClient uses the API-M-method
 \texttt{modifyDatastreamByReference} - otherwise \texttt{addDatastream}. 
The call of the Fedora-API-M-method \texttt{modifyDatastreamByReference} is executed by the following command:
\begin{verbatim}
call.invoke(new Object[] { pid, // the PID
  "ARTICLE", // name of the data stream
  null, // altIds (no alternative identifier)
  null, // dsLabel (no change - still "ESCIPUB")
  new Boolean(true), // versioning on
  mimeType, // the MIME type identified
  creator, // formatURI (author)
  localURL, // dsLocation (local URL)
  "A", // dsState (state - active)
  "update", // logMessage (log entry)
  new Boolean(true) }); // force
\end{verbatim}
Note that our prototype saves the name of the author in the object attribute \texttt{FormatURI}, from where this information can be retrieved
more easily than by analyzing the Dublin Core-data stream.
The article- and comment-lists, that are displayed to authors while reworking their articles
and to quality assurers while reviewing articles, make use of this information.

After deleting the article file in the Tomcat-root-directory, the TaskArticleBean determines the next transition in method \texttt{close}.
Then the task is finished and the user-interface layer returns to the working environment of the author.

At the lower end of the screen (cf. figure \ref{fig:screenshot}) the process graph is displayed; the current task is surrounded by a red rectangle.
The process graph is created by the jBPM class \texttt{ProcessImageTag.java}.
This class creates an HTML-table and adds the \texttt{.jpg}-image\footnote{This image was created by the \emph{Graphical Process Designer} and handed over to the jBPM
data base during deployment (cf.~section~\ref{sec:JBossjBPM}).} of the process graph as a background image of the table.
By evaluating the information in file \texttt{gpd.xml}\footnote{This file was created by the GPD and contains information about position and size of the elements of the process graph.},
the cells in the table get their red surrounding frames. The HTML-table is returned to the calling JSP.
The page \texttt{task\_author.jsp} is one of the JSPs of the prototype that uses this functionality. The call is made by the tag\\
\verb|<jbpm:processimage task="\${taskArticleBean.taskInstanceId}" />|

\subsection{Deployment of Process Definitions}
The deployment of process definitions using the Eclipse-plugin is only possible if \emph{eSciPub} is activated on the server.
\emph{eSciPub} contains an upload-servlet, acting as an interface between the deployment-functionality of the Eclipse-GPD-Plugin and the \emph{jBPM}-server.

If the deployment-function is called from the environment of the process administrator, the deployment does not use this servlet, but executes the corresponding jBPM-function directly.
The implementation of this functionality will be described now.

As a first step, the process administrator chooses a process archive file using a file selection dialogue of his workspace.
When he presses the ``Activate'' button,
an upload is performed using the Tomahawk-JSF-extension in the method \texttt{deploy} of the AdminBean.
The selected process archive file is transformed into a byte stream by\\
\verb|ByteArrayInputStream bais = |\\
\verb|      new ByteArrayInputStream(_upFile.getBytes());|\\
Since the process archives from GPD are compressed zip-files, the byte stream is decompressed by an instance of  \texttt{java.util.zip.ZipInputStream}:\\
\verb|ZipInputStream zipInputStream = new ZipInputStream(bais);|\\
Using method \texttt{parseParZipInputStream} of class\\
 \texttt{org.jbpm.graph.def.ProcessDefinition}, 
the process definition of the archive can be assigned to a new \texttt{ProcessDefinition}-instance and then be deployed by method deployProcessDefinition:
\begin{verbatim}
ProcessDefinition processDefinition =
  ProcessDefinition.parseParZipInputStream(zipInputStream);
jbpmContext.deployProcessDefinition(processDefinition);
\end{verbatim}
The deployment of the process archive is now finished.

\section{Discussion}
Although this work has been motivated by a scientific context, the concepts are general enough to be used by
any organization that needs to manage content for internal or external purposes.

We have provided a proof of concept for the integration of an open-source digital repository into a state-of-the-art enterprise architecture.

The existence of a graphical user interface for the configuration of workflows may give the impression that it is easy for administrators to configure those workflows. The good news is, that adding, removing and modifying the order of tasks and the deployment of the resulting new workflows is a matter of a few minutes. The graphical representations of the workflows, that show the status of a workflow instance to the end user, are provided by the system. The bad news is, that the resulting workflow is not necessarily sound (i.\,e.\ functioning) and that no error messages are displayed during creation and deployment. At present, the only way to find out whether the new workflow is operational is testing.
The documentation of a set of tasks should contain pre- and post-conditions that specify which tasks can be predecessors or successors of others. 
I.\,e.\ one should find out whether a static analysis can be added to jBPM that checks the soundness of workflows.

The jBPM Starters Kit\footnote{s.~\url{http://www.jboss.com/products/jbpm/downloads}} is a supplement to jBPM offered by JBoss. 
It contains a small web application, which provides tasks with very simple, string-based input fields. 
Based on this web application, our prototype implements custom-made tasks containing file handling and communication to Fedora.
In the current prototype, there is no straightforward way of combining custom-made tasks with standard tasks.
One direction to which future work should move, is the creation of a set of standard custom-made tasks that can be combined with jBPM standard tasks.
Another more complex direction might be the extension of the expressivity of jBPM itself.
\bibliographystyle{abbrv}
\def\btxandlong{and}
\bibliography{literatur}

\end{document}